\documentclass[twocolumn,11pt]{article}

\usepackage{titlesec}

\usepackage{flushend}

\usepackage{graphicx}

\titleformat{\section}
  {\normalfont\large\bfseries}
  {\thesection}{.5em}{}
  
\titleformat{\subsection}
  {\normalfont\bfseries}
  {\thesubsection}{.5em}{}
   
  \titleformat{\paragraph}[runin]
  {\normalfont\bfseries}
  {\theparagraph}{0pt}{}

\usepackage[margin=0.97in,top=.99in,bottom=1.1in, textheight=22.51cm,textwidth=16.51cm, heightrounded,columnsep=0.81cm]{geometry}

\usepackage{amsfonts,amsmath}

\usepackage{mathptmx}
\usepackage{times}

\usepackage{siunitx}

\pdfoutput=1  

\usepackage{cite,url}

\usepackage[font={scriptsize}]{caption}
\usepackage[font={scriptsize}]{subcaption}

\usepackage[colorlinks,bookmarksopen,bookmarksnumbered,citecolor=red,urlcolor=red]{hyperref}

\newlength{\noteWidth}
\long\def\notes#1{\ifinner
           {\footnotesize #1}
           \else
           \marginpar{\parbox[t]{\noteWidth}{\raggedright\footnotesize #1}}
       \fi\typeout{#1}}

\def\notes#1{\typeout{read notes: #1}}  

\def\Proof{\paragraph{Proof}}

\def\UA{U^{\text{\sf\tiny A}}}
\def\bfmUA{\bfmath{U}^{\text{\sf\tiny A}}}

\def\sq{\hbox{\rlap{$\sqcap$}$\sqcup$}}
\def\qed{\ifmmode\sq\else{\unskip\nobreak\hfil
\penalty50\hskip1em\null\nobreak\hfil\sq
\parfillskip=0pt\finalhyphendemerits=0\endgraf}\fi\medskip}

\long\def\defbox#1{\framebox[.9\hsize][c]{\parbox{.85\hsize}{%
\parindent=0pt
\baselineskip=12pt plus .1pt      
\parskip=6pt plus 1.5pt minus 1pt 
 #1}}}

\long\def\beginbox#1\endbox{\subsection*{}%
\hbox{\hspace{.05\hsize}\defbox{\medskip#1\bigskip}}%
\subsection*{}}

\def\endbox{}

\newsavebox{\junk}
\savebox{\junk}[1.6mm]{\hbox{$|\!|\!|$}}

\def\bfmath#1{{\mathchoice{\mbox{\boldmath$#1$}}%
{\mbox{\boldmath$#1$}}%
{\mbox{\boldmath$\scriptstyle#1$}}%
{\mbox{\boldmath$\scriptscriptstyle#1$}}}}

\def\bfmD{\bfmath{D}}

\def\bfmU{\bfmath{U}}

\def\bfmY{\bfmath{Y}}

\def\bfmhhaY{\bfmath{\hhaY}} 
\def\bfmhhaY{\hbox to 0pt{$\widehat{\bfmY}$\hss}\widehat{\phantom{\raise 1.25pt\hbox{$\bfmY$}}}}

\def\til={{\widetilde =}}

 \def\FRAC#1#2#3{\genfrac{}{}{}{#1}{#2}{#3}}

\def\ddtp{{\mathchoice{\FRAC{1}{d^{\hbox to 2pt{\rm\tiny +\hss}}}{dt}}%
{\FRAC{1}{d^{\hbox to 2pt{\rm\tiny +\hss}}}{dt}}%
{\FRAC{3}{d^{\hbox to 2pt{\rm\tiny +\hss}}}{dt}}%
{\FRAC{3}{d^{\hbox to 2pt{\rm\tiny +\hss}}}{dt}}}}

\def\eqdef{\mathbin{:=}}

\def\average#1,#2,{{1\over #2} \sum_{#1}^{#2}}

\def\eye(#1){{\bf(#1)}\quad}

\newtheorem{theorem}{Theorem}[section]

\newtheorem{lemma}[theorem]{Lemma}

\def\Section#1{Section~\ref{#1}}

\newcounter{rmnum}
\newenvironment{romannum}{\begin{list}{{\upshape (\roman{rmnum})}}{\usecounter{rmnum}
\setlength{\leftmargin}{14pt}
\setlength{\rightmargin}{8pt}
\setlength{\itemsep}{2pt}
\setlength{\itemindent}{5pt}
}}{\end{list}}

\newcounter{anum}

\def\Ebox#1#2{%
\begin{center}
\includegraphics[width= #1\hsize]{#2} \end{center}}

\def\Fig#1{Fig.~\ref{#1}}

\title{\vspace{-1cm}
Smart Fridge / Dumb Grid?
\\
Demand Dispatch for the Power Grid of 2020
}

 \author{\normalsize
 \centering
 \begin{tabular}{ccccccc} 
 Joel Mathias & \hspace{0.15cm} &	
 Rim Kaddah & \hspace{0.15cm} & 
 Ana Bu\v{s}i\'c &  \hspace{0.15cm} &	
 Sean Meyn 
 \\ 
Univ.~of Florida &&
Telecom Paristech 	&&
Inria/ENS 	&&
 Univ.~of Florida 
 \\
 \small
 joel.mathias@ufl.edu &&	
  \small
 rim.kaddah@telecom-paristech.fr &&	
 \small	
 ana.busic@inria.fr  && 
 \small   
 meyn@ece.ufl.edu 
   \end{tabular}   
   }
   
      \date{}

\begin{document}

\maketitle

\bigskip

\bigskip

\paragraph{Abstract} 
In discussions at the 2015 HICSS meeting, it was argued that loads can provide most of the ancillary services required today and in the future. Through load-level and grid-level control design, high-quality ancillary service for the grid is obtained without impacting quality of service delivered to the consumer. This approach to grid regulation is called \textit{demand dispatch}: loads are providing service continuously and automatically, without consumer interference. 

 In this paper we ask, \textit{what intelligence is required at the grid-level}? In particular, does the grid-operator require more than one-way communication to the loads? Our main conclusion: risk is not great in lower frequency ranges, e.g., PJM's RegA or BPA's balancing reserves. In particular, ancillary services from refrigerators and pool-pumps can be obtained successfully with only one-way communication. This requires intelligence at the loads, and much less intelligence at the grid level.

\smallskip
\noindent
 \textbf{Acknowledgements}
 {\it Research supported by the NSF grant CPS-1259040, and the French National Research Agency grant ANR-12-MONU-0019.}

\section{Introduction}
\label{s:intro}

Recent FERC rules require that ISO/RTOs provide incentives to responsive resources to help balance supply and demand in the power 
grid. FERC Order 755 directs RTOs and ISOs to  provide a ``payment for  performance that reflects the quantity of frequency regulation service provided by a resource when the resource is accurately following the dispatch signal''.
Preliminary studies suggest that these incentives are having desirable impact.  In this paper, we investigate the value of performance.   

\begin{figure}[!h]
\Ebox{1}{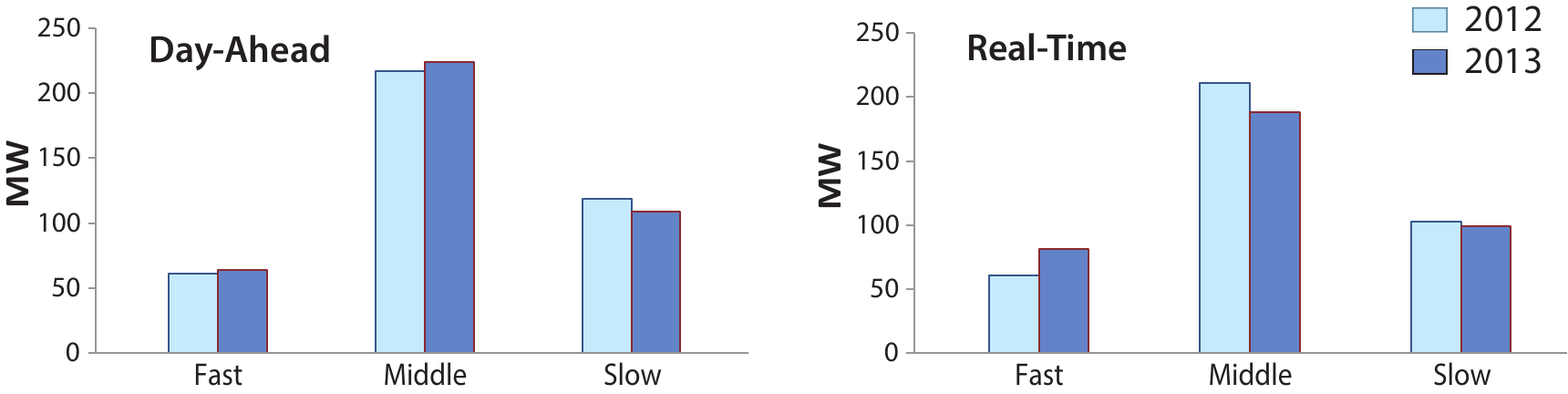}
 \vspace{-1em}
\caption{Average total daily regulation volume at MISO:
Availability of fast-ramping resources expanded after MISO introduced payment for performance.} 
\vspace{-.3em}
\label{f:miso}
\end{figure}

One year after adoption of  new   performance incentives at MISO,  the availability of fast ancillary services increased significantly in the real-time market.  The report  \cite{MISO13} on which 
\Fig{f:miso} is based also demonstrates improved control performance since the adoption of these incentives.


New market rules for regulation services  adopted at PJM were also successful.  It was  found that their overall costs were reduced, even with the higher payments to ancillary service providers \cite{xiasubrefrecarschola14}.  

Pages 23 and 24 of the FERC 755 report \cite{FERC755} 
contain a survey of experiments conducted by Beacon Power and Primus Power on the value of highly responsive resources for ancillary service.  Primus claims that this results in approximately 76 percent more ACE correction (compared with what  can be obtained from generation sources).  This is because a ``\textit{slower ramping resource lags to the point of working against needed ACE correction}''.

%

\notes{"in spite of" is directly interchangeable with despite.
}

These empirical results show that high-performance may add value.  On the other hand, the grid has been remarkably reliable despite poor performance from traditional sources of ancillary service.  Kirby in \cite{kir04} shows that regulation obtained from coal generators is in some cases excellent, and in other cases the generators show significant delay in response.  There is no evidence that poor tracking performance led to grid outages,  but it is likely that there are hidden costs.

One goal of the research surveyed in this paper is to explain why the grid is so resilient to the disturbances introduced to the grid through the poor regulation performance described by Kirby,   and why it is nevertheless sensible to incentivize high performance for certain types of regulation services.  

The ultimate goal of this research is to create needed ancillary services through ``intelligent loads''.  In prior research surveyed at the 2015 HICSS meeting, it was argued that loads can provide most of the  ancillary services required today and in the future \cite{barbusmey14}.  
To ensure reliability to grid operators and to consumers,  ancillary services from loads should be provided continuously and automatically, without consumer interference. This approach is henceforth called \textit{demand dispatch} \cite{bro10}.     

A key element in prior work is a frequency decomposition of the regulation signal,  and a classification of loads
based on the bandwidth of ancillary service they can provide
\cite{haomidbarey12,linbarmeymid15,meybarbusyueehr14,barbusmey14}.
Provided intelligence at each load is designed so that these and other constraints are recognized,   
high-quality ancillary service for the grid is obtained without impacting quality of service delivered to the consumer.

\begin{figure}[!h]
\Ebox{1}{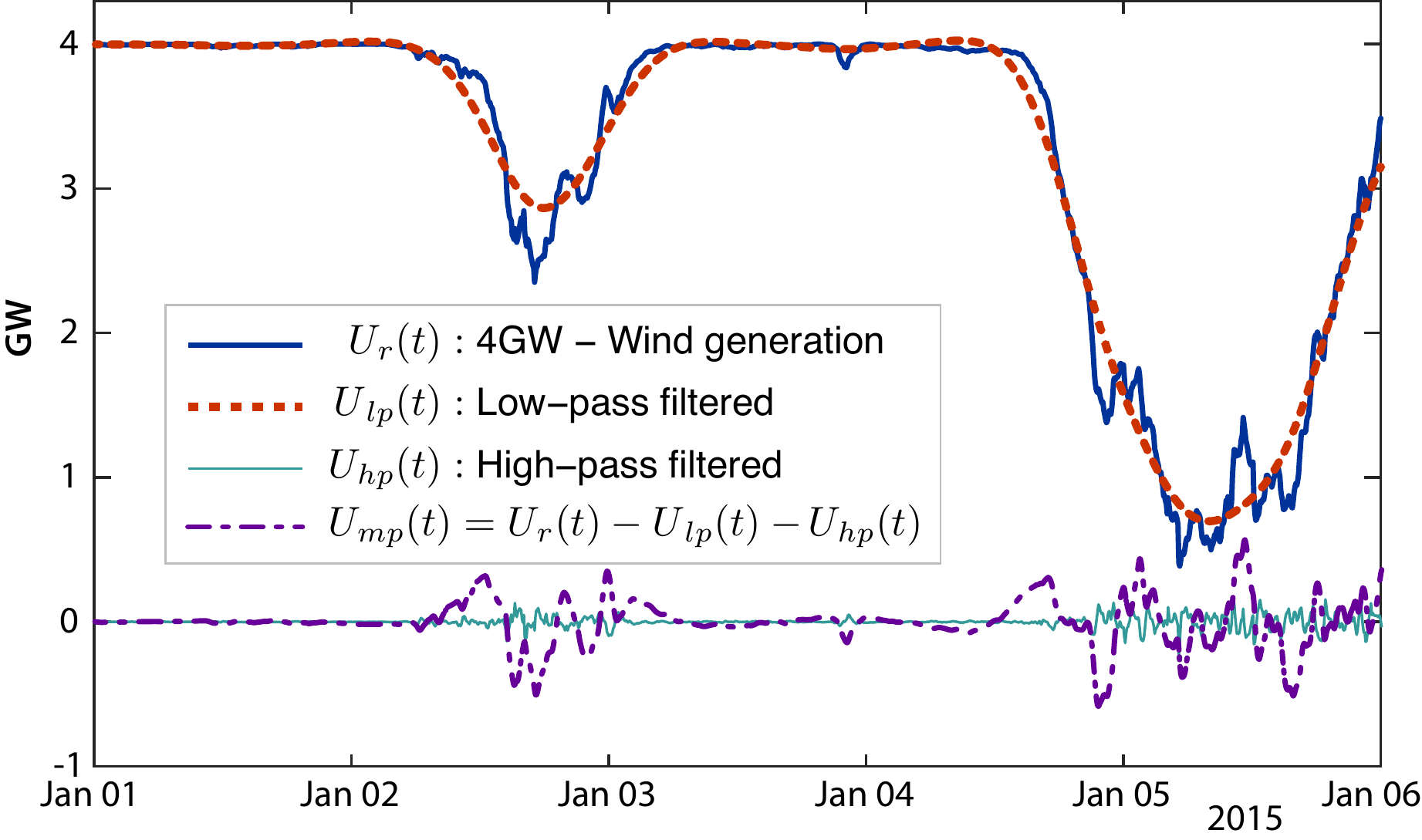}
\vspace{-1em}
\caption{Decomposition of residual generation $U_r$ into three products, differentiated by frequency content, 
to serve a 4~GW load at BPA.}
\vspace{-.25em}
\label{f:BPA3}
\end{figure}
To illustrate the application of these concepts,  consider the Bonneville Power Authority (BPA) region in the Northwestern United States during the first week of January, 2015 \cite{BPA}.   As an exercise, suppose that a constant 4~GW load must be served, in spite  of volatile power from wind.   The power that is not provided by wind generation is denoted by $U_r(t)$ in \Fig{f:BPA3}.  Observe that $U_r(t)\approx 4$ on New Years Day since there was little power from wind on that day.  At other times, power from wind generation is nearly 4~GW, so that $U_r(t)\approx 0$.

The signal $\bfmU_{lp}$ is obtained by passing $\bfmU_r$ through a \textit{non-causal} low-pass filter.  The justification is that
the day-ahead forecast of the low-frequency component of the wind is very accurate. 
$\bfmU_{hp}$ is obtained by passing $\bfmU_r-\bfmU_{lp}$ through a causal high-pass filter,
and $\bfmU_{mp}$ is the remainder of needed power. 

The time-scale of $\bfmU_{hp}$ is similar to the valuable RegD signal used at PJM \cite{pjm-man15}.    In research conducted at the University of Florida, it is shown that loads in commercial HVAC systems can provide this ancillary service with negligible cost, and  no impact on building climate \cite{haomidbarey12,linbarmeymid15}.  The time-scale of $\bfmU_{mp}$ corresponds to what can be provided from thermostatically controlled loads (TCLs), as considered in  \cite{johThesis12}, residential pool pumps \cite{barbusmey14}  (a load of approximately 1GW in Florida or California), and many other loads.  

The low frequency deviation $\bfmU_{lp}$ can be obtained from generation sources such as thermal or hydro.  
Some or all of this can   be obtained from ramping up and down power consumption in aluminum manufacturing or other large commercial loads.  

In this paper we ask,  \textit{what intelligence is required at the grid-level to implement demand dispatch}?   In particular,  does the grid-operator require more than one-way communication to the loads?   The question is investigated using two separate lenses.  The first is in terms of system-wide risk:  does poor performance threaten grid stability?  The second set of questions are framed in terms of system-wide cost.  In particular,  if 50\%\ of resources provide poor-quality ancillary service,  what is the impact on the other 50\%?

%
 \begin{figure*}
\Ebox{.75}{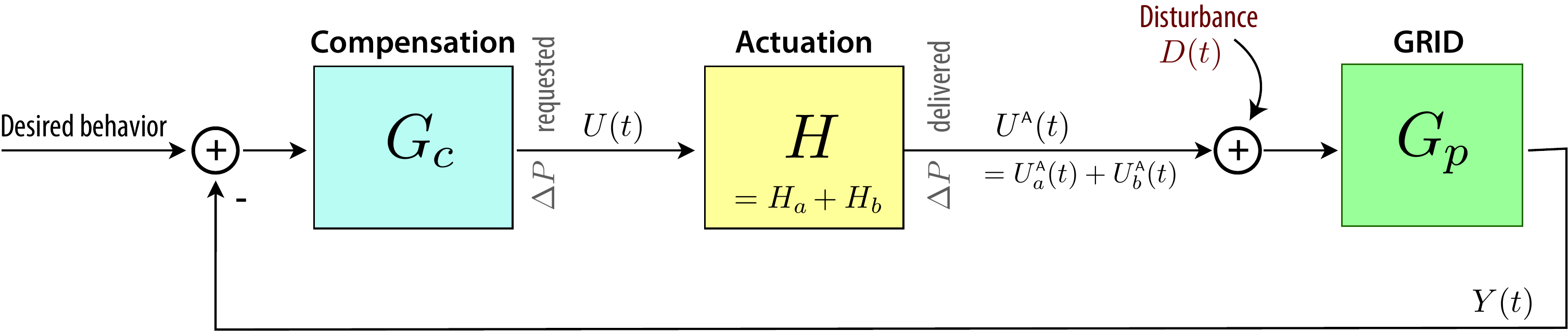}  
\vspace{-1em}
\caption{Power Grid Control Loop.   A question addressed in this paper:  \textit{where do we find $H$?}}
\label{f:CAv2}
\vspace{-.75em}
\end{figure*}

The analysis is based on a standard input-output model of the grid \cite{kun94}, denoted $G_p$ in \Fig{f:CAv2}.  The ``desired behavior'' will be zero, and $Y(t)$ will represent a deviation whose desired value is zero.   The actuation block in the figure is obtained using demand dispatch along with conventional sources of ancillary service.  

Using arguments from classical control, it is argued that it makes sense to pay a premium for accurate regulating reserves in higher frequency bands,  while the value of performance is not so great in lower frequency bands:

 \begin{romannum}

\item 
The highest bandwidth of ancillary service considered in this paper is on time-scales corresponding to primary reserves.  This is the region of highest risk,  which is not surprising to grid operators who are conservative in setting droop parameters for governors.

\item
Risk is not great in lower frequency ranges;  time-scales ranging from several minutes to hours.

At these lower frequency ranges,  there is little cost for poor performance, but there may be high cost for heterogeneity.  In particular, if 50\%\ of resources provide poor quality ancillary service in terms of phase lag,  then the other 50\%\ may be forced to provide  greater service to the grid.     

It is argued that the impact of heterogeneity can be reduced through an additional layer of local control at each load.

\end{romannum}
The good news:  two-way communication may not be required.  If intelligence at the loads is designed appropriately,  then the grid will obtain the required regulation with only one-way communication from balancing authority to these intelligent loads.  

The greatest benefit of demand dispatch is for slower time-scales such as the balancing reserves at BPA~\cite{BPA},  or PJM's RegA signal \cite{pjm-man15}.   We can obtain all of these sources of ancillary service from flexible loads with no risk to the grid,  and no loss of service to consumers.

Of course, even if there is no cost, the consumer needs incentives to participate.  We envision engagement through contracts, much like the Florida OnCall program.  It is likely that there will be a fixed payment for engagement, and regular payments that are proportional to ``services rendered'' (much like the mileage payments in place today, in response to FERC order 755 \cite{xiasubrefrecarschola14}).

\paragraph{Related research}

Models that can capture the dynamics of the entire grid have been extensively studied, 
and have received attention recently  due to concerns about grid inertia. The recent work \cite{chaghaeri14} studies the impact of new trends, such as the deployment of inverter interfaced generation on the dynamics of the grid.

 


The paper \cite{peybal12}, following \cite{chabalsha12},  shows the diverse effects that fast resources can have on system stability. These grid models are used in our investigations here.



Dynamic models of major household appliances and their DR potential is 
investigated in \cite{malcho88,johThesis12} (see also the references therein).

Controlling loads using a randomized policy was studied in several works, such as \cite{johThesis12,meybarbusyueehr14}. The masters thesis \cite{broThesis09} proposes  randomization to solve the synchronization problem that arises in application of Schweppe's FAPER algorithm for distributed frequency control \cite{schFAPER80}. 

 \medskip
 
 The remainder of this paper contains three additional sections, organized as follows:
 \Section{s:control} contains the details of the three components of the block diagram shown in
 \Fig{f:CAv2}.   The grid model $G_p$ and compensator $G_c$ are based on standard and recent sources.    Greater attention is devoted to the actuator block,  which includes  grid-level models for a combination of several classes of loads involved in demand dispatch.  In \Section{s:smart}, control concepts are applied to a specific scenario involving just two classes of loads working in conjunction with traditional ancillary service resources.  Simulations illustrate the main conclusions. Conclusions and directions for future research are contained in \Section{s:conc}.

\section{A Distributed Control Problem}
\label{s:control}

The realization of the three power products shown in \Fig{f:BPA3} can be obtained using a hierarchical control strategy.   At the grid level,  the low frequency signal $\bfmU_{lp}$ could be obtained from a day-ahead market or contracts with generators and commercial loads.  

The higher frequency components of $\bfmU$ can be obtained using demand dispatch.
 The `intelligence' at each load involves local control at each device to achieve two goals:  reliable service to the grid, and high quality of service to the customer that the device serves.    The  main goal of this paper is to understand the level of reliability required.

We begin with a model of the grid.

\subsection{Grid dynamics}
\label{s:Gp}

A transfer function model of the grid is denoted $G_p$ in \Fig{f:CAv2}.  
We assume here that the input to $G_p$ is power deviation, and the output is the deviation of grid frequency from its nominal value  (60~Hz in the U.S.).     
In practice, we would also measure tie-line error and other grid disturbances.

 An approximate grid model can be obtained by considering an interconnection of aggregate generation and governor dynamics, as in \cite{chabalsha12} and other papers.  The final model is justified by observing the dynamics of the grid following a fault;  the initial trajectory can be interpreted as the transient portion of a step response.  

  \begin{figure}[h!]
\Ebox{.85}{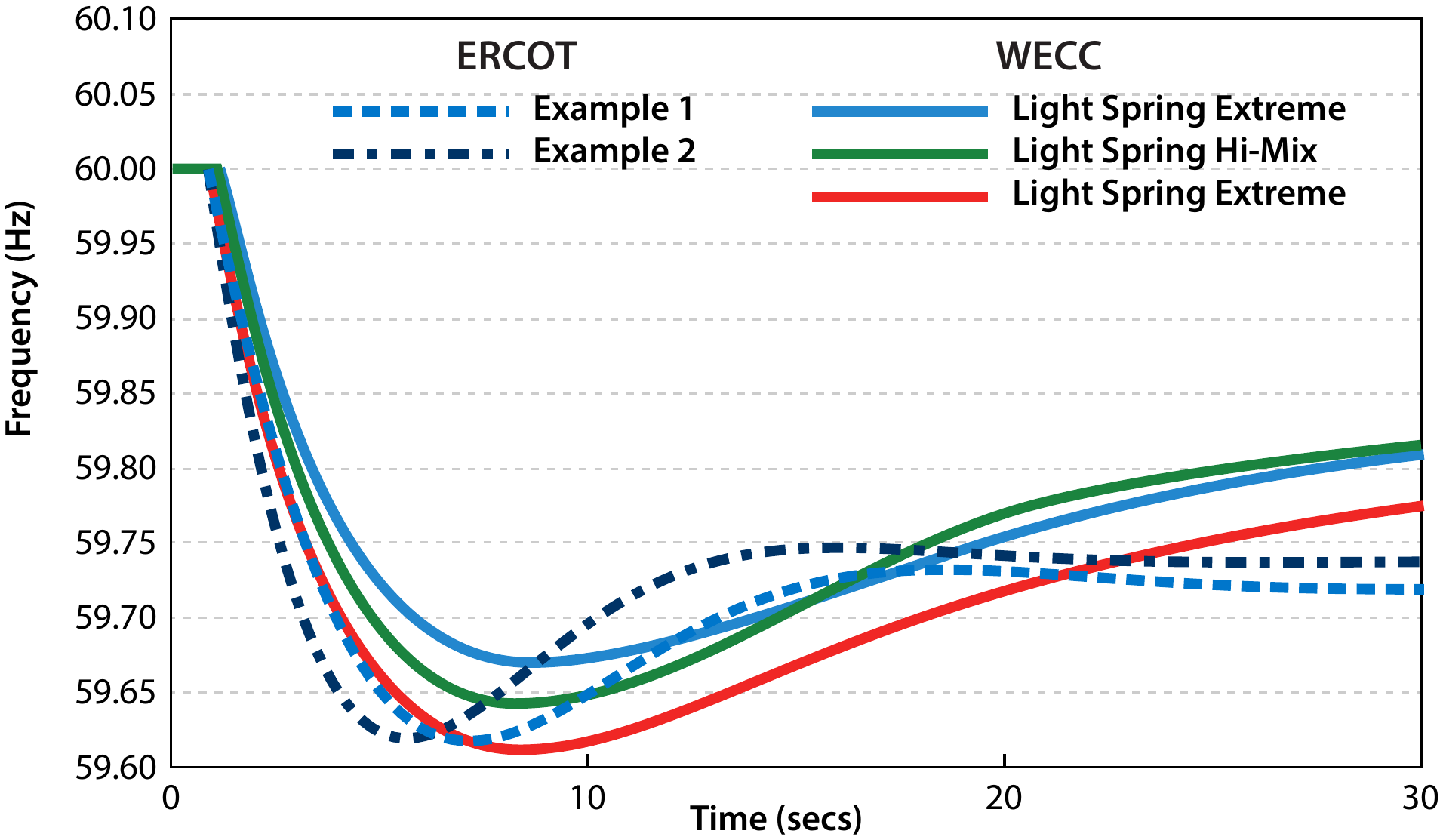}
 \vspace{-1em}
\caption{Frequency deviations following a generation outage -- examples from WECC and ERCOT.}
\label{f:WECC-ERCOT}
\end{figure}
 
\Fig{f:WECC-ERCOT} shows examples from WECC (Western US; taken from 
Fig.~57 of \cite{milshapajdaq14}) and ERCOT (Texas)  \cite{peybal12}.  
The slow recovery of grid frequency to its nominal value in the WECC plots is due to secondary control.  This is not a part of $G_p$ since we do not consider contingency reserves in this paper.   

In theoretical models of the grid,  the transfer function depends on the nominal load.   
This is consistent with observations of the grid following a fault \cite{kun94,chabalsha12}.  
Consequently, there is significant model uncertainty  and variability that must be respected in control design at the grid level.

One version of the ERCOT model is used in the simulation studies summarized in \Section{s:smart}:  the grid model of \cite{chabalsha12} was used in which the net load 
 (load minus wind generation)
 is $25$~GW, resulting in
\begin{equation}
G_p(s) =
   \frac{ 0.644 s + 0.147}
   {
 s^2 + 0.4797 s + 0.147}
\label{e:ERCOTex1}
\end{equation}
Corresponding grid frequency dynamics are shown in Example~1 of \Fig{f:WECC-ERCOT}.  This transfer function has natural frequency $\omega_n= 0.3834$,  and damping ratio $\zeta =  0.6256$.     The large frequency excursion seen in \Fig{f:WECC-ERCOT} is due to the zero in $G_p$.

\subsection{Actuator dynamics}
\label{s:act}

The transfer function $H$ will be the sum of several transfer functions whose dynamics are shaped by the dynamics of the resource providing ancillary service  and some pre-filtering --- both locally and at the grid-level.  Consider the case of PJM in which $H_a$ and $H_b$ can be associated with  RegD and RegA, respectively.   Because accurate tracking is required for RegD,   the transfer function $H_a$ is shaped entirely at PJM,
and the signal $ \UA_a(t)$ will accurately track a multiple of the RegD signal.
The transfer function $H_b$  will include a model of the  dynamics of the resources providing the service.

Based on industry practice, and to simplify control design, the following convention is imposed on the feedback architecture:  the signal $\bfmU$ in \Fig{f:CAv2} is scalar valued. Subsequently, it may be decomposed into the sum of several signals differentiated by bandwidth, but these dynamics are modeled in $H$.  
 
In our prior work using two-way communication combined with local control,  the loads could track the desired regulation signal nearly perfectly.    Without two-way communication,  the loads will introduce both dynamics and uncertainty in the transfer function $H$.

   \begin{figure}[h]
 \Ebox{1}{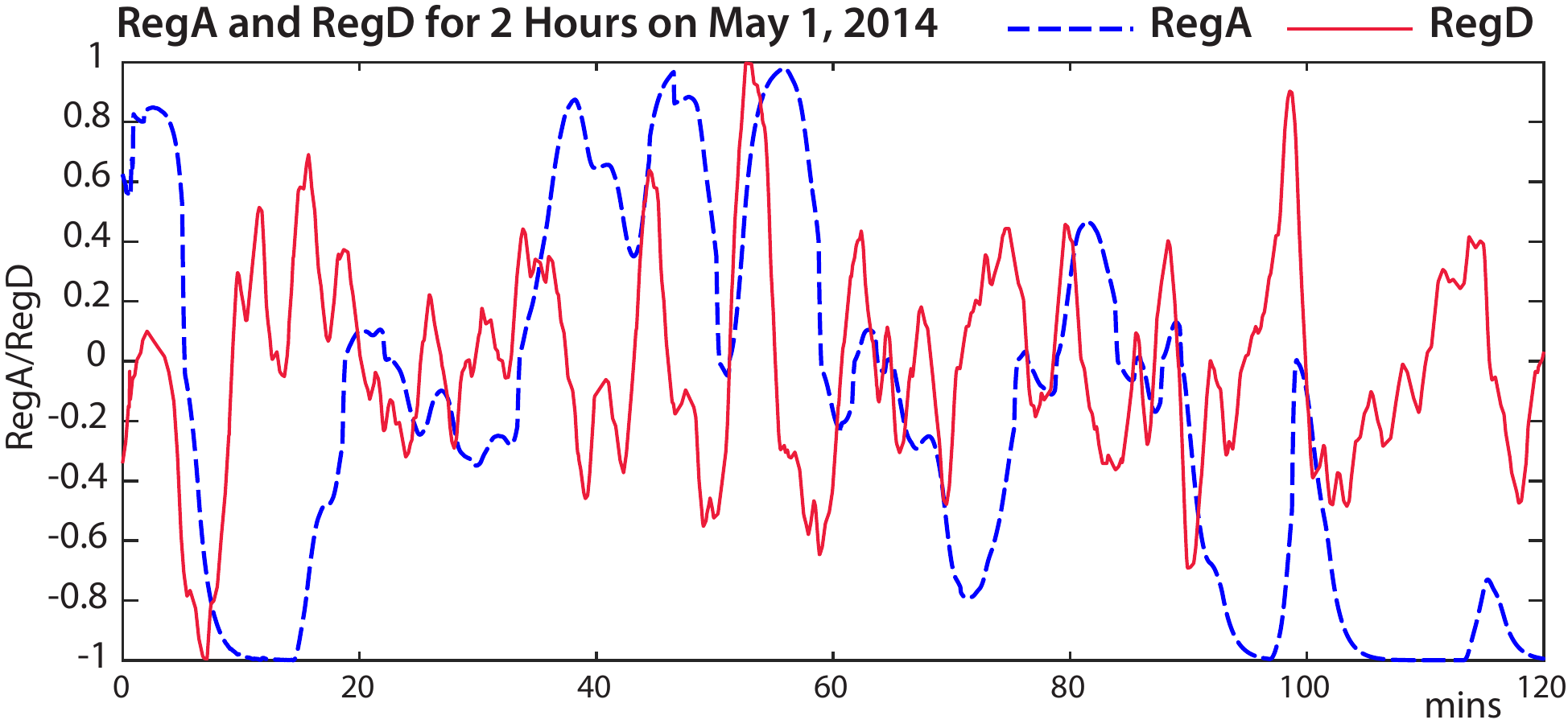}
\vspace{-1em}
\caption{RegA and RegD at PJM for two hours in May, 2014.}
\label{f:RegAD2hr}
\end{figure}

We are, however, free to introduce additional local control to improve grid-level performance.     \Section{s:smart} contains examples in which load dynamics exhibit resonance and some phase-lag. Let $G_l$ denote a transfer function that models the dynamics for a particular class of loads.  These dynamics are easily identified, so additional filtering at the load is used to improve the input-output behavior.
The dynamics of the filtered loads are expressed 
\begin{equation}
\label{e:Hl}
H_l(s) = M_l(s) G_l(s) 
\end{equation}
The pre-filter $M_l$ is designed so that the collection of loads has reduced resonance, improved phase response, and greater bandwidth. 

This design step has not appeared in prior work.   The overall design is illustrated through examples in \Section{s:smart}.


\subsection{Grid disturbance}
\label{s:GridDist}

Before turning to the design of the compensator $G_c$, we first consider how PJM defines the signal $\bfmU$ appearing in \Fig{f:CAv2}.  It is obtained by first constructing the area control error (ACE), which is a linear combination of grid frequency deviation and tie-line error.  The ACE signal is passed through a PI compensator (their $G_c$) \cite{pjm-man15}, and this is then transformed into a sum of two signals,  RegA and RegD.   \Fig{f:RegAD2hr} compares the two signals over a two hour time period.  The higher frequency content in RegD is evident from the figure.


 \begin{figure}[h]
 \Ebox{1}{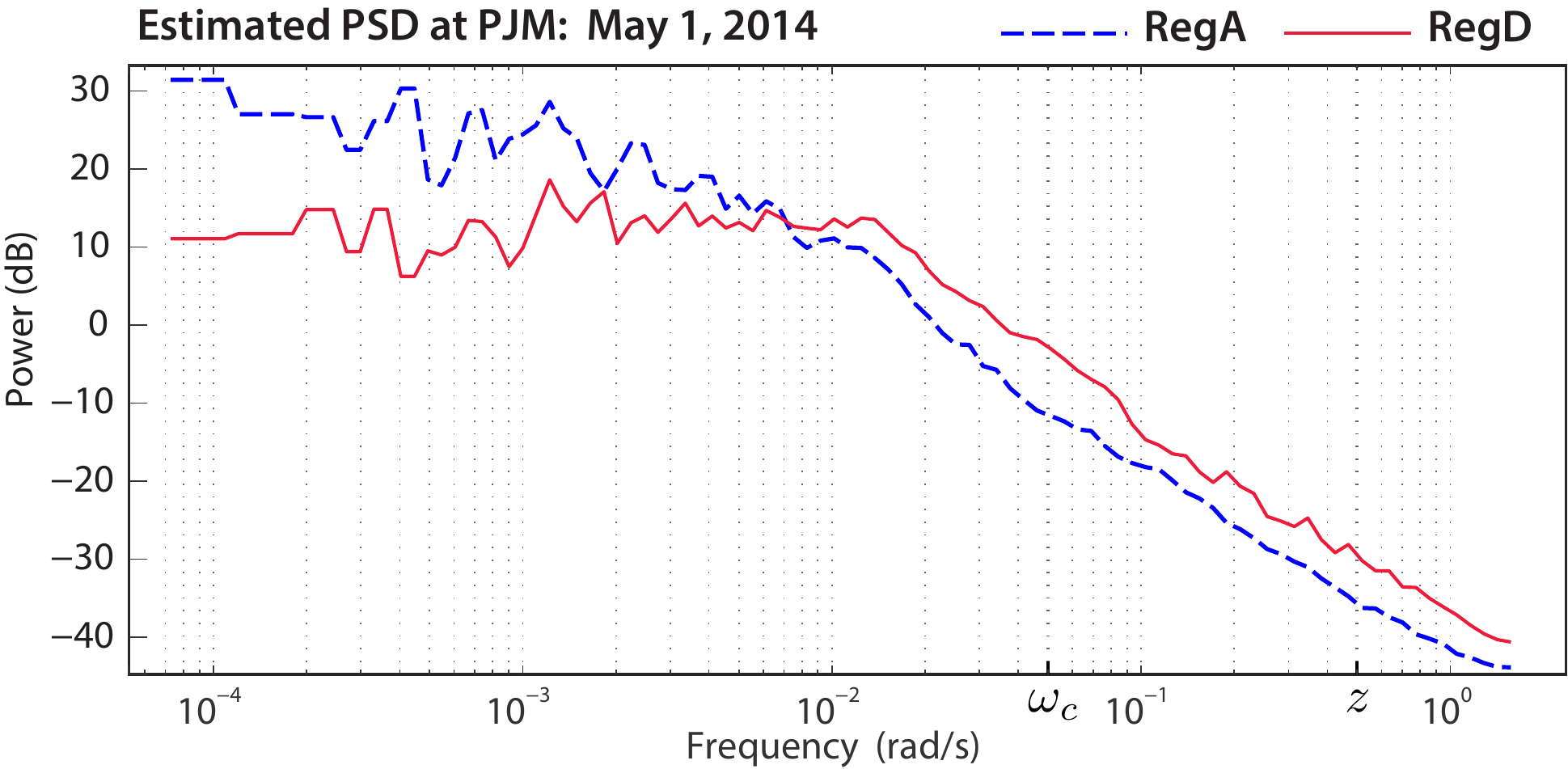}
\vspace{-1em}
\caption{PSD for the RegA and RegD at PJM based on 24 hours of data.}
\label{f:psdAD2hr}
\end{figure}

Power spectral density (PSD) estimates were obtained using time series over a 24 hour period on the same day.  A comparison of the plots shown in \Fig{f:psdAD2hr} shows again the higher frequency content in RegD.  It also shows very little energy in frequencies greater than $10^{-1}$~rad/s.  

It follows that a compensator $G_c$ at PJM will require high gain for frequencies as high as about $10^{-2}$,    and relatively low gain at frequencies above $\omega = 10^{-1}$~rad/s.

The balancing reserves at BPA have significantly greater low frequency content.  At ISO/RTOs,  this low frequency regulation is obtained in the real-time market.  


\subsection{Control design}
\label{s:controldesign}

The \textit{loop transfer function} associated with the feedback loop in \Fig{f:CAv2} is the product of the three transfer functions:
$L(s) = G_c(s) H(s) G_p(s)$.   The crossover frequency is   by definition the value for which $|L(j\omega_c)|=1$.
This frequency is unique by design,  and chosen by the grid operator through the choice of parameters in the   compensator $G_c$.   The grid transfer function $G_p$ is approximated by a second-order transfer function, whose natural frequency $\omega_n$ depends on load.
\notes{depends on which grid:
 ranging from about $0.2 $ to $1$~rad/s.  }
Because of this uncertainty,  the compensator should suppress the gain of the loop transfer function in this range of uncertainty ($\omega_n$ ranges from $0.2 $ to $1$~rad/s in the ERCOT model).

Following standard design for power systems and elsewhere, we adopt the  PI compensator,
\begin{equation}
G_c(s) = K \frac{s+\beta}{s}
\label{e:PI}
\end{equation}
In the simulations described later in the paper, 
we take $\beta=0.5$ and fix the gain $K$ so that $\omega_c=0.05$~rad/s; the crossover frequency $\omega_c$ is known to approximate the closed loop bandwidth. 
With this design we hope to achieve two goals: disturbance rejection in the frequency range where disturbances are present, and robustness to uncertainty in grid dynamics.

\subsection{Cost of heterogeneity}
\label{s:hetero}

Suppose that we have two sources of ancillary service:   
the first provides perfectly accurate service, but is costly.
The second is free, but inaccurate.   Our control solutions are insensitive to un-modeled dynamics, especially at low
frequency.  We may ask, does the introduction of the poor-quality service create a cost, because the more expensive services must work harder?

To address this question we introduce two transfer functions $G_a$ and $G_b$, modeling accurate and inaccurate resources, and for $\rho\in[0,1]$ denote
\begin{equation}
H=H_a+H_b = (1-\rho) G_a + \rho G_b
\label{e:HGaGb}
\end{equation}
In this subsection we are concerned with the magnitude of 
service delivered by the accurate actuators,
denoted $\UA_a(t)$ in \Fig{f:CAv2}; 
in transfer function notation,  $\UA_a/U =  (1-\rho) G_a $.
The steady-state analysis is based on the following representation:

\begin{lemma}
\label{t:UaD1}
The transfer functions from the disturbance to output, and from disturbance to $\UA_a$ are given by, respectively,
\begin{eqnarray}
\frac{Y}{D}  &=&  \frac{G_p}{1+ L}
\label{e:yd1}
\\[.2cm]
\frac{\UA_a}{D} &=&  - (1-\rho) \frac{L_a}{1+ L}
\label{e:UaD1}
\end{eqnarray}
in which $L=G_c H G_p$ is the  loop transfer function,  and
 $L_a=G_c G_a G_p$.
\end{lemma}

\Proof
The first identity 
 is obtained using standard arguments:  
 from the figure we
 obtain $Y=G_p(D - G_c HY)$, since ``desired behavior'' is zero by definition.  Solving this algebraic equation gives \eqref{e:yd1}.

The following transfer function  is easily identified from the figure and the definition of $\UA_a(t)$:
\begin{equation}
\frac{\UA_a}{Y} =  - (1-\rho) G_a G_c 
\label{Ua1}
\end{equation}
The transfer function \eqref{e:UaD1}
is obtained on combining \eqref{e:yd1} and \eqref{Ua1}.
\qed

Consider a steady-state setting in which the disturbance $\bfmD$, shown in \Fig{f:CAv2}, is purely periodic:  $D(t) =\sin(\omega_0 t)$ for some $\omega_0>0$,
and all other signals are periodic with the same frequency.   
Applying \eqref{e:UaD1},  we must have $\UA_a(t) = k_0 \sin(\omega_0 t + \phi_0)$, 
where   $k_0=| \UA_a/D(j\omega_0) |$ and $\phi_0=\angle \UA_a/D(j\omega_0) $.
The gain $k_0 = k_0(\rho,\omega_0)$ is interpreted as the cost;  this is motivated by current mileage payments at ISOs \cite{xiasubrefrecarschola14}.

Normalize the transfer functions so that $|G_a(j\omega_0)| = | G_b(j\omega_0)| =1$.  
The first resource is assumed to be perfect,   $G_a(s)\equiv 1$, and we denote $\phi_B = \angle G_b(j\omega_0) $.

\begin{figure}
\Ebox{.95}{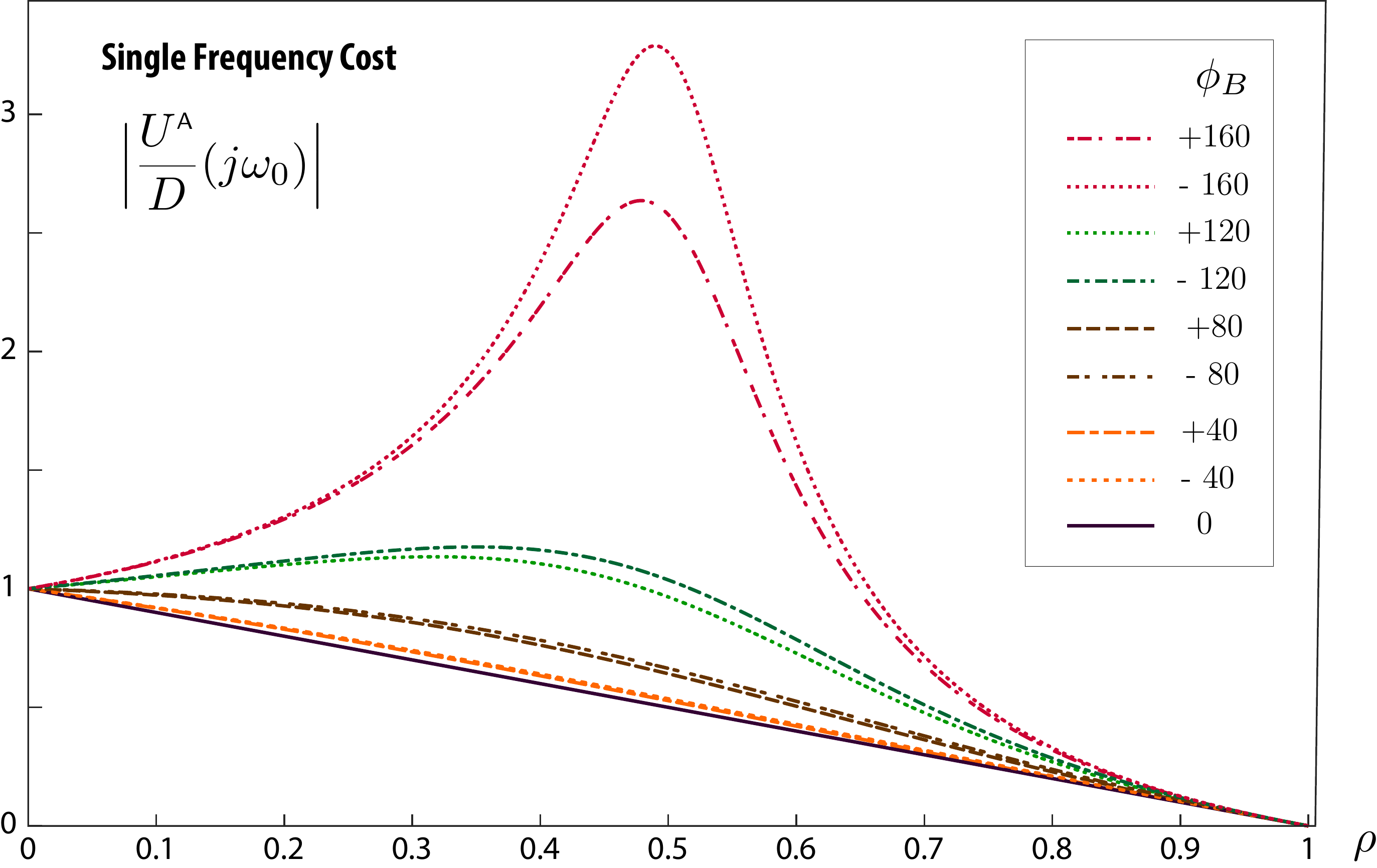}
 \vspace{-1em}
\caption{System cost at a single frequency:  
the highest cost is
expected with a mix of approximately $50\%$ of the two sources of ancillary service. } 
\vspace{-.3em}
\label{f:ZaDSFA}
\end{figure}

\Fig{f:ZaDSFA} shows the cost $k_0 $ 
as a function of $\rho$ for a range of values of $\phi_B$,  and a fixed   value of $\omega_0$.  
When there is no heterogeneity ($\phi_B=0$), 
the cost decays linearly.  For $\phi_B>90 $ degrees,  the cost is maximized near $\rho = 1/2$.   

The common slope at $\rho=1$ can be identified as,
\[
\frac{d}{d\rho}k_0(\rho,\omega_0) \Big|_{\rho=1}
=
-
\Bigl|\frac{L_a(j\omega_0) }{1+ L(j\omega_0) } \Bigr| 
\]

In conclusion:  
a free resource may be costly if it does not accurately follow the regulation signal.  However, the common slope at $\rho=1$ means that cheap resources are valuable,  provided they aren't mixed with others.  

All of these conclusions presume that the system is stable.  In the next section we consider both stability and cost in a more realistic model.

\begin{figure*}[t]
    \centering
\begin{subfigure}[t]{0.465\textwidth}
\Ebox{1}{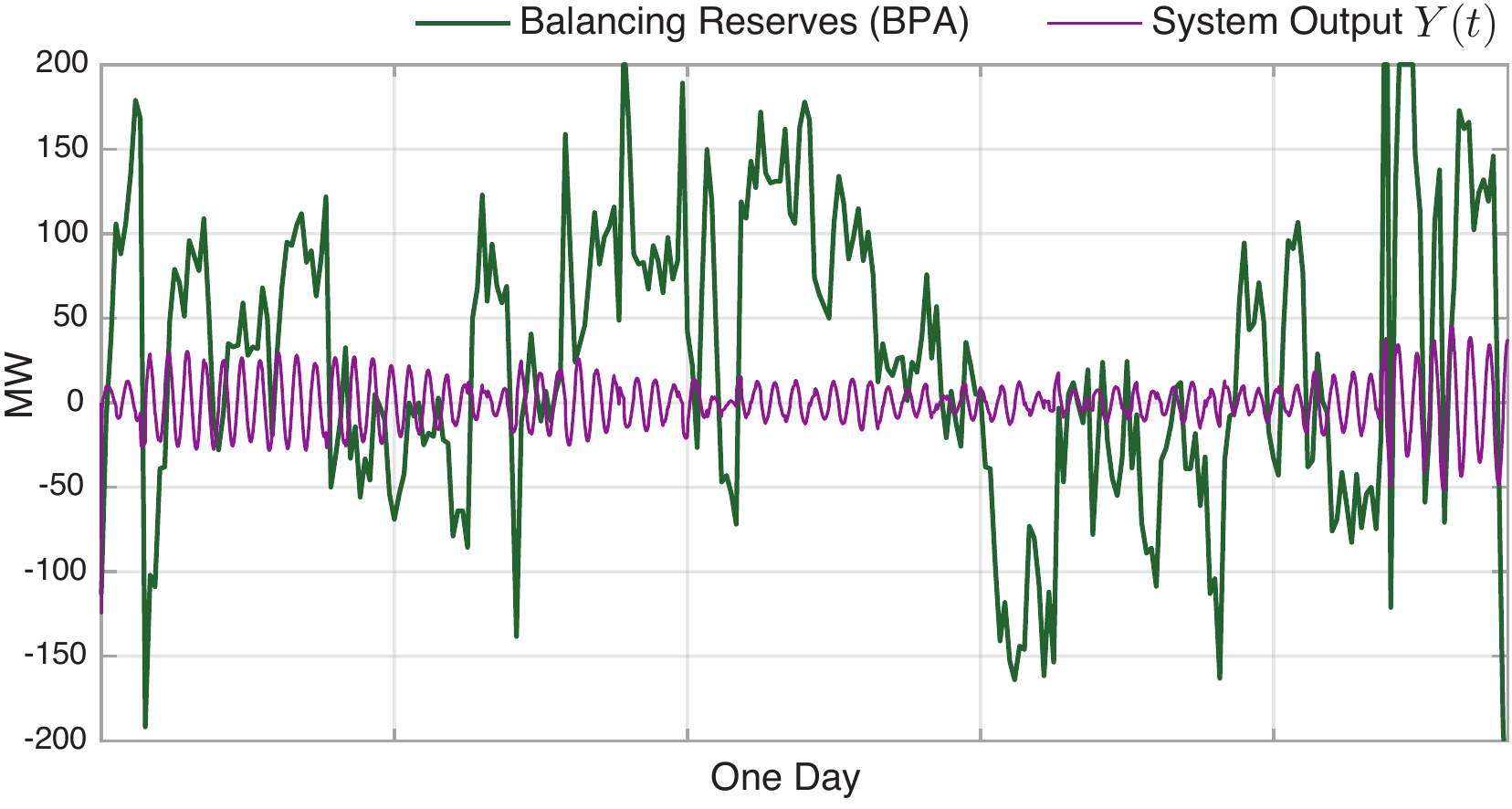}
\vspace{-1em}
\caption{Lead design with     $\omega_{co} = 0.007$~rad/s.  } 
\label{f:TrackingLead007}
\end{subfigure}
\hfill
 \begin{subfigure}[t]{0.465\textwidth}
\Ebox{1}{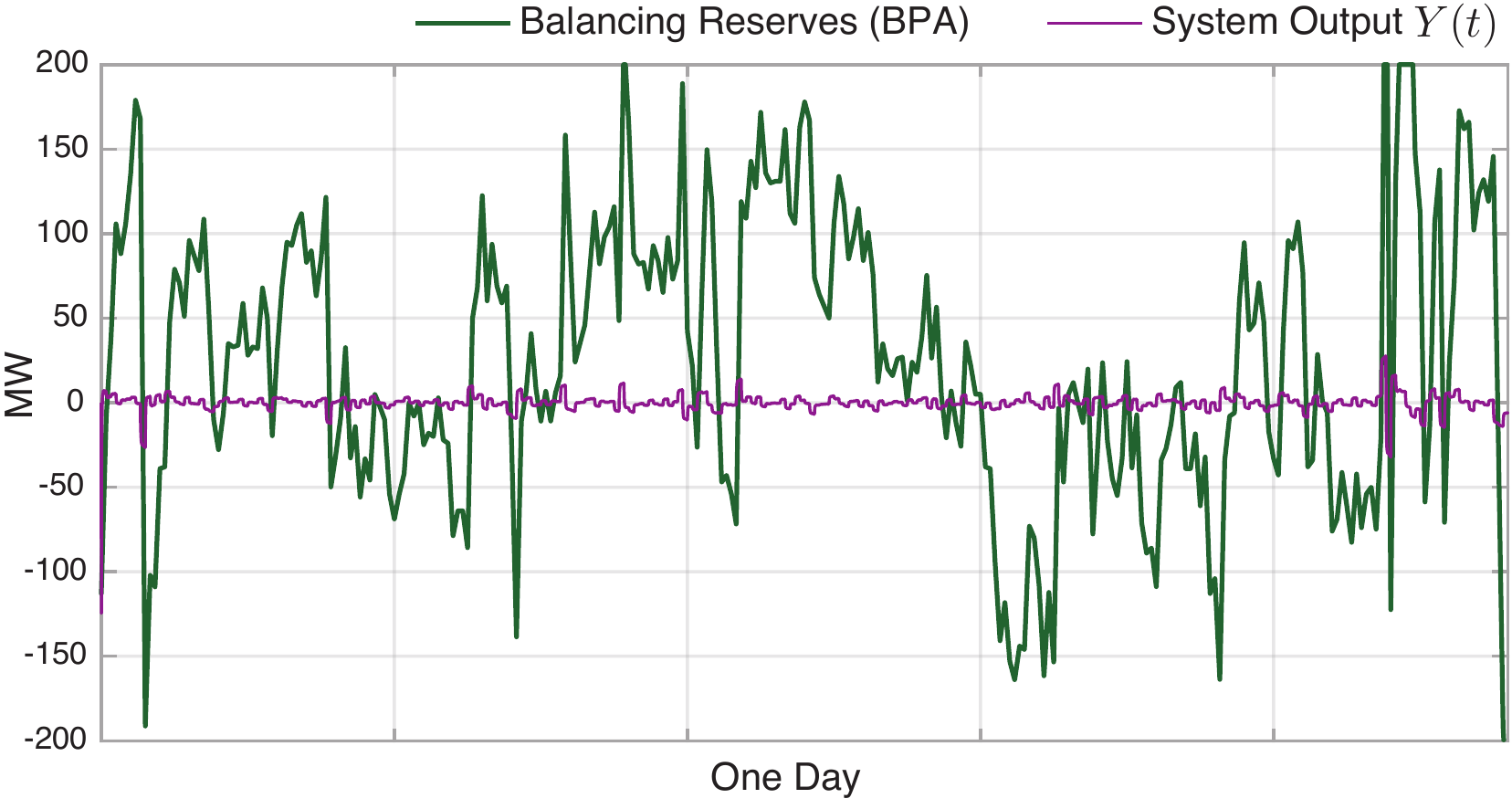}
\vspace{-1em}
\caption{Inverse design, with $\omega_{co} = 0.013$~rad/s.  }  
\label{f:TrackingInverseFilter013} 
\end{subfigure}
\vspace{-.2em}
    \caption{Comparison of disturbance rejection performance obtained with two different pre-filter designs.}
     \label{f:tracking}
\end{figure*}

\section{Risk \&\ Cost using Demand Dispatch}
\label{s:smart}

Experiments were conducted using demand dispatch, with one-way communication from grid to loads.

The grid transfer function $G_p$  in \Fig{f:CAv2} was taken to be the second-order linear system \eqref{e:ERCOTex1}, whose natural frequency is $\omega_n= 0.3834$.  As discussed in the \Section{s:Gp},  the grid transfer function is sensitive to load and generation mix.  This uncertainty means that we should maintain a closed loop bandwidth significantly below the natural frequency of the grid.  The PI compensator \eqref{e:PI} is designed to set the crossover frequency to $\omega_c = 0.05$~rad/s.

The disturbance $\bfmD$  was defined as the balancing reserves deployed at BPA during a typical week:  
June 6, 2015 to June 12, 2015 (data available at \cite{BPA}).       \Fig{f:tracking} shows a plot of $\bfmD$ during the first day of this week,
along with the closed loop response $\bfmY$ obtained using two different designs.  The poor performance seen in \Fig{f:TrackingLead007} is eliminated with better local control at the loads. These results are explained in \Section{s:track}. 

The actuator transfer function $H$ is described next.

\subsection{Actuator block}
\label{s:actBlock}

The actuator block $H$ in \Fig{f:CAv2} is obtained from a combination of three resources:
 a transfer function $G_a$ representing an expensive, high-quality resource, and  two actuators $ H_{pl} $  and $ H_{tcl}$, which are based on two separate collection of loads.   The two classes of loads work in parallel:  the `pools' could represent a large collection of residential pool loads, or other loads that provide flexibility in a low frequency range.  The `TCLs' represent flexible loads providing ancillary service in a higher frequency range;  these may include refrigerators, water heaters, and air-conditioners.  
 

Just as in \Section{s:hetero},  the performance of $G_a$  is assumed to be perfect,  
with $G_a\equiv 1$.

In addition,  we introduce a parameter $\rho$,  
a low pass filter $ H_{LP}$,  
a high pass filter $ H_{HP}$, 
and define the overall 
actuator transfer function as follows:
\begin{equation}
\begin{aligned}
H &=   H_{LP} H_\ell + (1-H_{LP}) G_a
\\
\text{\it where} \quad
H_\ell &= H_{pl} +   
H_{HP}  [ (1-\rho) G_a + \rho H_{tcl} ] 
\end{aligned}
\label{e:H}
\end{equation}
This design choice is based on several considerations:
\begin{romannum}
\item
The low pass filter $H_{LP} $ is introduced because of risk of instability due to gain or phase uncertainty from load response:
it  is
 designed so that the response from loads is not significant near the crossover frequency $\omega_c$. 
\item
The high-pass filter $H_{HP} $ is used to limit the low-frequency content of the regulation signal sent to the TCLs.  This is to help guarantee QoS (quality of service) constraints for these loads.
\item
The purpose  of the parameter $\rho$ is to investigate cost as a function of the proportion of TCLs engaged.  
It will be seen that their imperfect response may introduce cost to the system.
\end{romannum}

For $H_{LP} $, we used a second-order Butterworth low pass filter with cut-off frequency 
$\omega_{co}$   and damping ratio $\zeta_{lp}=\sqrt{2}/2$, with transfer function $H_{LP}(s) = \omega_{co}^2/(s^2 + 2\zeta_{lp} \omega_{co}s + \omega_{co}^2)$.  The cut-off frequency $\omega_{co}$ is a parameter used in our cost/risk evaluations.  

Various values for the cut-off frequency were considered, but  the upper bound $\omega_{co} \le 0.013$~rad/s was imposed throughout.  This is not because the loads cannot provide service at higher frequencies, but because the response from uncertain loads should be attenuated near the crossover frequency $\omega_c=0.05$~rad/s.

For the high-pass filter $H_{HP} $, a second-order Butterworth  filter was used, with cut-off frequency of 0.0004 rad/s:
$H_{HP} (s) =  
              s^2/(   s^2 + 0.0005657 s + \num{1.6e-07})$.   This was chosen based on the bandwidth constraints for pools, and QoS constraints for TCLs.


Further details are provided in the next subsection.

\textit{In the numerical experiments reported here, we have simulated the linear mean-field model of \cite{meybarbusyueehr14}, and not the collection of loads}.
In all  prior work it is found that the mean-field deterministic model reflects actual behavior very accurately, especially in a control setting.  This is true even when only 100 loads are engaged \cite{chebusmey15}.  A more detailed simulation taking into account transmission and perhaps distribution will be the subject of future work.  

\subsection{Load models and pre-filter design}
\label{s:mm}

The transfer function $H$ in \Fig{f:CAv2} is obtained from a combination of resources, including demand dispatch.
In prior work it has been shown that a randomized control architecture for each load leads to a tractable input-output model for the aggregate mean-field model  \cite{barbusmey14}.  
It is nonlinear,  but a linearized model worked well for purposes of estimation, control and performance evaluation 
\cite{chebusmey14,chebusmey15}.  In the case of residential pools pumps, the focus of these three papers,  the transfer function for the linear model has a strong resonance at a frequency corresponding to a period of 24 hours.   The transfer function depends on the number of hours of cleaning per day, but the resonance is independent of this parameter. 
A second-order approximation was adopted,
\begin{equation}
G_{pl} =\frac{\omega_{pl}^2}{s^2 + 2\zeta_{pl} \omega_{pl} s +\omega_{pl}^2},
\label{e:Gpl}
\end{equation}
with natural frequency $\omega_{pl} = \num{7.27e-5}$~rad/s; this 
corresponds to the 24-hour periodic behavior of pools.
\notes{not exactly, but ok}

The behavior of a typical TCL is similar to a pool filtration system. 
Take, for example, a typical residential refrigerator.  Its  nominal behavior is similar to a pool pump, with two exceptions.  First,  the behavior is roughly periodic with period much shorter than 24 hours.  Second,  in many cases the operating time is a smaller fraction of this period.
We have obtained an input-output model using a technique similar to what is introduced in  \cite{barbusmey14}, and find that a collection of residential refrigerators with a 30 minute cycle-time admits a linear system model with resonance corresponding to this period.  Accordingly, a second-order approximation for the aggregate of TCLs was used in these experiments,
\begin{equation}
G_{tcl} =\frac{\omega_{tcl}^2}{s^2 + 2\zeta_{tcl} \omega_{tcl} s +\omega_{tcl}^2},
\label{e:Gtcl}
\end{equation}
with natural frequency $\omega_{tcl} = 0.003$~rad/s, which approximately corresponds to the 30-minute cycle-time.

The damping ratios were  set  to  $\zeta_{pl} = \zeta_{tcl} = 0.5$.   Other values were tested with similar results.   

To obtain the two transfer functions $H_{pl} $ and $ H_{tcl} $ appearing in
\eqref{e:H}
requires one additional ingredient.

In prior work,
 it is shown that 
the bandwidth of ancillary service provided by loads can be extended through feedback.  Even though the natural frequency for the linear model of \cite{meybarbusyueehr14} corresponds to 24 hours,  a grid-level control solution extends the closed loop bandwidth   by one decade,  resulting in a closed-loop natural frequency of approximately
$  \num{7.27 e-4}$~rad/s, and without resonance.  


In the experiments described here, we do not allow communication from load to grid, so we cannot extend bandwidth using the approach of  \cite{meybarbusyueehr14}.   Instead, the bandwidth is set through the design of the pre-filter introduced in  \Section{s:act}.

The local filters used at the pools and the TCLs are denoted $M_{pl}$ and $M_{tcl}$, respectively. Following \eqref{e:Hl}, we obtain the two actuator transfer functions,
\begin{equation}
\label{e:HpoolHtcl}
\begin{aligned}
H_{pl} = M_{pl} G_{pl}
 \qquad
H_{tcl} = M_{tcl} G_{tcl}
\end{aligned}
\end{equation}
The overall actuator transfer function of \eqref{e:H} is written,
\begin{equation} 
H =  H_a+H_b\eqdef P_a G_a + H_{LP} H_B 
\label{e:Hb}
\end{equation}
in which $H_B=[ H_{pl} +   \rho H_{HP} H_{tcl} ] $ and
\begin{equation}
\label{e:Pa}
P_a = 1 - H_{LP} [ 1 - H_{HP} (1 - \rho)]
\end{equation} 
When $\rho =1$,  the transfer function $H$ is a convex combination of good and potentially ``bad'' actuators: 
\[
H =     (1-H_{LP} ) G_a +H_{LP} H_B 
\] 

Two designs were considered for the local filter.
\begin{romannum}
\item \textit{Lead design:} 
For $\tau > 0$ and $\alpha < 1$,
\begin{equation}
M_l = \Bigl(\frac{\tau s+ 1}{\alpha\tau s +1}\Bigr)^2
\label{e:lead}
\end{equation}
\item \textit{Inverse design:}
For $\alpha < 1$,
\begin{equation}
\label{e:inv1}
M_{l} = (\alpha + G_{l})^{-1}
\end{equation}
\end{romannum}

The inverse design is intended to 
approximate the inverse of the transfer function for the  aggregate load model.  In very recent work,  we have found that a better approximation can be obtained using concepts from robust control theory --- this topic will be explored in detail in future work.

The lead design   
\eqref{e:lead} 
 consists of two lead compensators in series: this is to counter the $-40$~dB/decade slope after the resonance in the load transfer functions. The lead compensator is a high-pass filter, so the bandwidth of the loads is increased by some value depending on $\alpha$ and $\tau$.
 
 The lead design was applied to both pools and TCLs: 
 we chose $\tau_{pl} = 15,000$ and $\tau_{tcl} = 350$, so that $\tau^{-1}$ is slightly smaller than the natural frequencies of the respective loads. In addition, we set $\alpha_{pl} = 1/15$ and $\alpha_{tcl} = 1/5$. Using these settings for the lead design, we achieve the required bandwidth extension for pools and TCLs.

The inverse design was also applied to both classes of loads.  In the simulations that follow we only show results obtained when the inverse design was used to construct $M_{tcl}$; it was found that  sensitivity of performance to the choice of the pre-filter  $M_{pl}$ is not great.
\notes{please check}

 \begin{figure}
\Ebox{1}{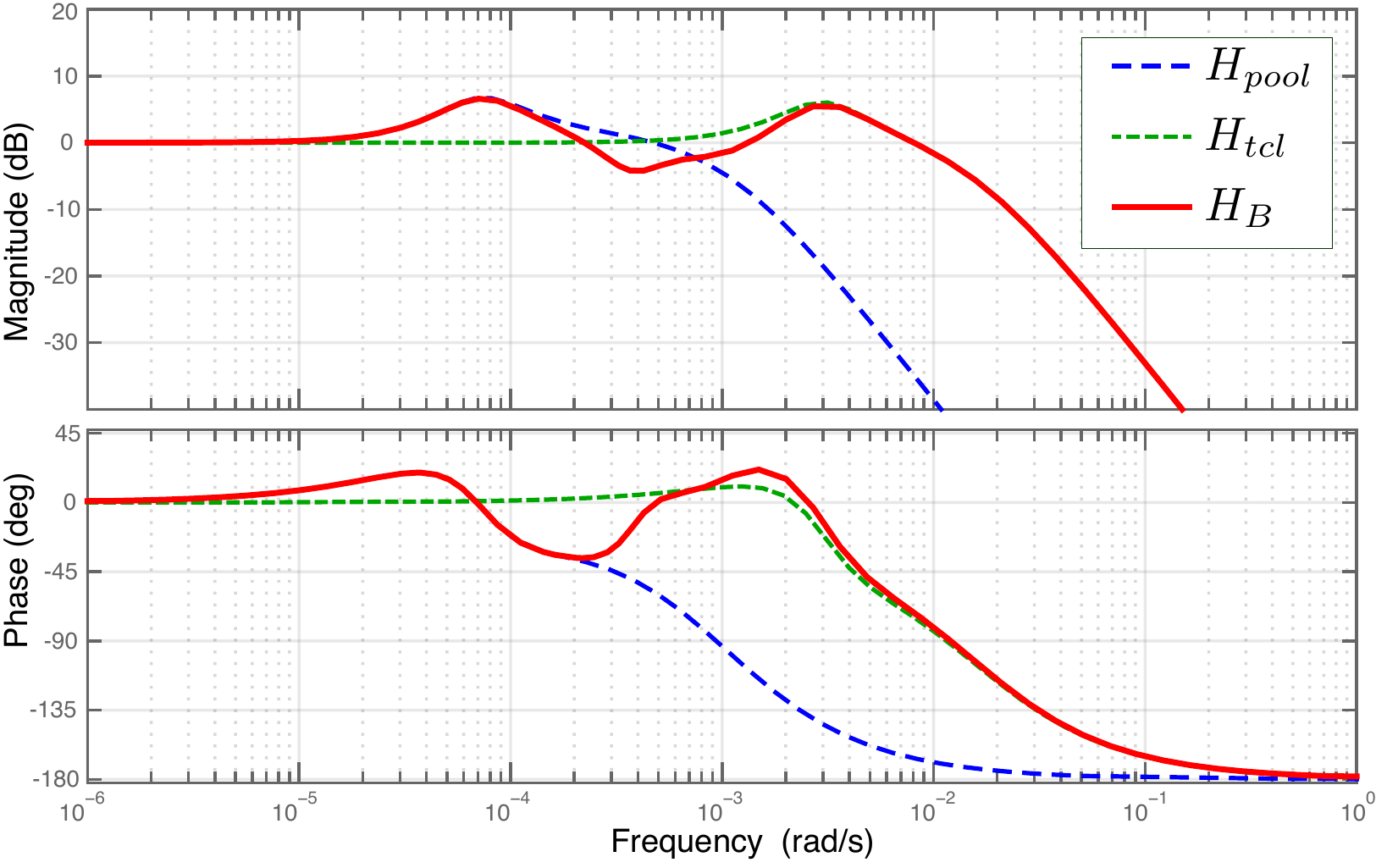}
 \vspace{-1em}
\caption{$H_B$ obtained using the lead  design.  } 
\vspace{-.3em}
\label{f:PoolTCLHbBode}
\end{figure}

\notes{so confusing!  How can we explain that we didn't use inverse for pools?  What was alpha for TCLs?  
\\
We also consider the inverse design for TCLs, represented in \eqref{e:inv1}: 
the TCL pre-filters its input by an . 
}

\Fig{f:PoolTCLHbBode} shows a Bode plot of $H_B$ using the lead design with $\rho=1$.
The rapid phase decline at high frequencies motivates the cut-off frequency at $\omega_{co} =0.003$~rad/s.    We do not observe large phase lag using the inverse design for the TCLs,  so we can take a larger value for $\omega_{co} $ in this case.

\notes{not very important:
The phase dip near $\num{2e-4}$ could be suppressed with a more careful pre-filtering of the regulation signal received by the pools.
}
 
We will look more closely at  the impact of these design choices in the following.

\subsection{Transfer functions to model risk \&\ performance}

Recall that the loop transfer function is $L = G_c H G_p$.  The corresponding \textit{system sensitivity transfer function} is defined in control texts as
\begin{equation}
\label{e:S}
S = \frac{1}{1+L},
\end{equation}
and its maximum gain is denoted
\begin{equation}
\label{e:Ms}
M_S = \| S \|_\infty = \sup_\omega\left|S(j\omega)\right|
\end{equation}

\begin{figure}[h]
\Ebox{1}{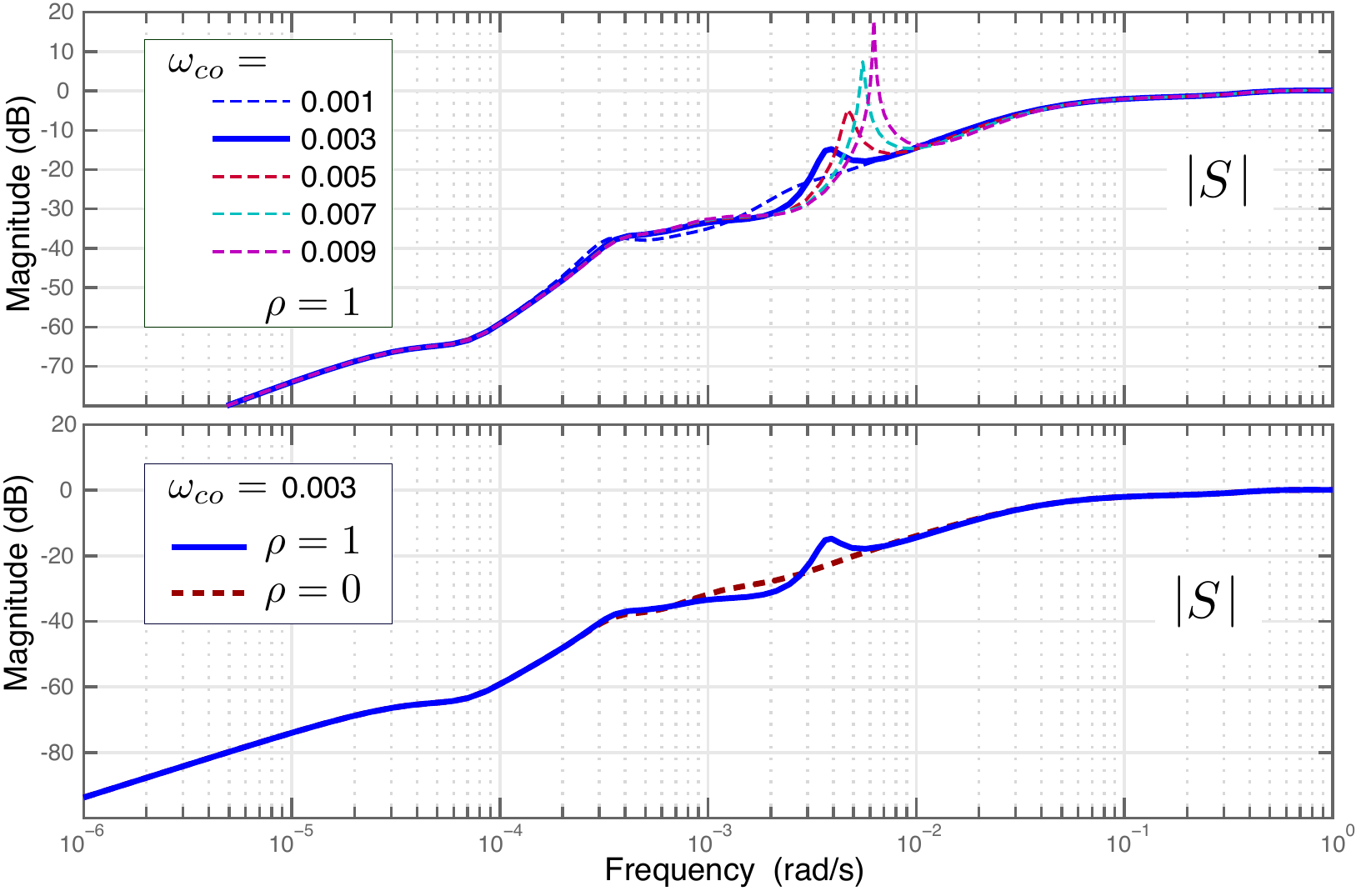}
 \vspace{-1em}
\caption{Sensitivity using the lead design.
The closed-loop system is not sensitive to the imperfect response from TCLs
when $\omega_{co} =0.005$ or smaller. } 
\vspace{-.3em}
\label{f:SensitivityBode+2rhos}
\end{figure}
 
The terminology is motivated by the interpretation,
\[
|S(j\omega)| = d(L(j\omega), -1)
\]
where $d(L(j\omega), -1) = |L(j\omega) + 1|$.  
The minimum $\min_\omega |L(j\omega) + 1|$ is called the \textit{vector  margin} or the \textit{stability margin}.   If this is zero, then there is a closed loop pole on the imaginary axis, so the system is not stable.  From the definitions, a small stability margin is equivalent to a large maximal sensitivity $M_S$.

\Fig{f:SensitivityBode+2rhos} shows the magnitude plot of the sensitivity
function \eqref{e:S}
 for various values of $\rho$ and $\omega_{co}$ using the lead  design.  The maximal sensitivity grows quickly with $\omega_{co}$ when $\rho=1$.  With $\omega_{co}=0.003$~rad/s, we find that the sensitivity plots are similar for $\rho=0$ or $\rho=1$.

 Sensitivity using the inverse design is lower for any $\omega_{co} \leq 0.013$~rad/s, and  
 the value of $\rho$ has little impact on sensitivity for this range of  $\omega_{co}$.

\begin{figure}[!h]
\Ebox{1}{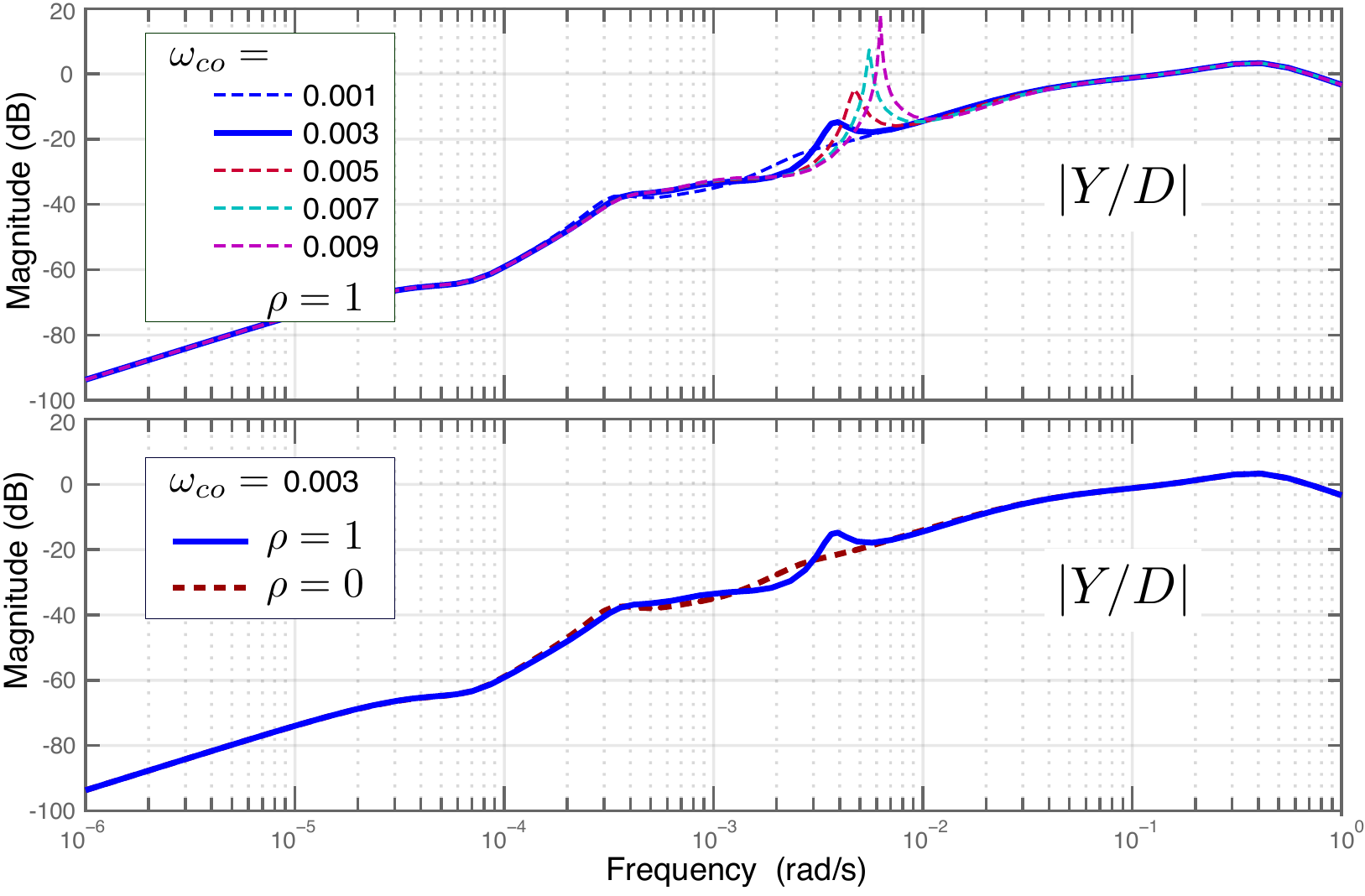}
 \vspace{-1em}
\caption{Disturbance rejection using the lead design.  } 
\vspace{-.3em}
\label{f:YD2rhos}
\end{figure}

Recall from \eqref{e:yd1} that  the transfer function from disturbance   to output   is  $Y/D = G_p/(1+ L)$.   \Fig{f:YD2rhos} shows Bode plots of this transfer function for the lead design:
Using   $\omega_{co} =0.003$~rad/s,
the system effectively suppresses disturbances whose frequency is below 
$\omega=10^{-2}$.   Performance degrades when using larger values of $\omega_{co}$.  
For the inverse design, the transfer function $Y/D$ is largely independent of $\rho$  for any  $\omega_{co}\le 0.013$~rad/s. 

 \begin{figure*}[t]
 \centering
\includegraphics[width= .95\hsize]{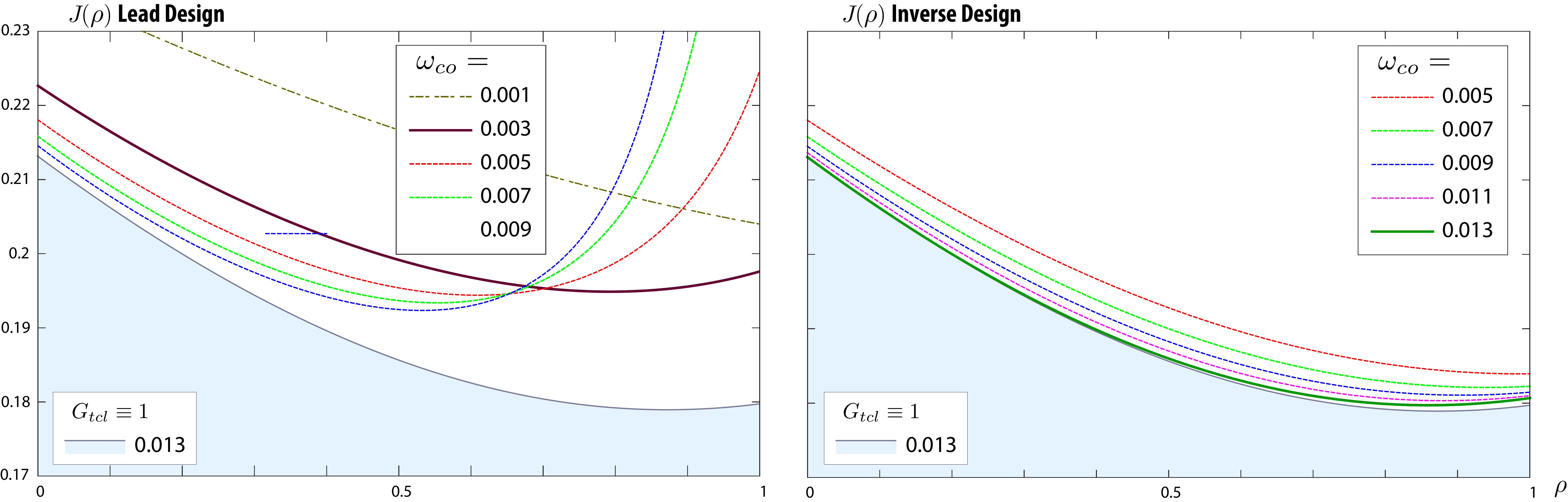}
 \vspace{-.5em}
\caption{Cost $J$    as a function of $\rho$ and $\omega_{co}$ for the two designs.   The cost obtained using the inverse design is similar to what is obtained using an ideal actuator,  $H_{tcl}\equiv 1$. } 
\vspace{-.3em}
\label{f:H2}
\end{figure*}

\subsection{Mean-square cost}

Let $\UA_a(t)$ denote the signal supplied by the high-quality actuators.   
Based on the block diagram shown in \Fig{f:CAv2} and the definition of $H$ in \eqref{e:H},
we have
\begin{equation}
\frac{\UA_a}{D} = -P_a\frac{L_a}{1+L},
\label{e:UaDsec3}
\end{equation}
in which  $L_a=G_c G_a G_p$.  The proof  of \eqref{e:UaDsec3}
 is similar to the derivation of \eqref{e:UaD1}.

The maximal sensitivity is denoted
\begin{equation}
M_a = \Bigl\| {\frac{\UA_a}{D}} \Bigr\|_\infty = \sup_\omega \Bigl|\frac{\UA_a}{D}(j\omega)\Bigr|
\label{e:Ma}
\end{equation}
While \eqref{e:Ma}
is   a common performance metric in the control literature,   
the maximizing frequency may have little to do with the frequencies encountered in operation.   

The following $L_2$ cost is based on a statistical model of the disturbances entering the grid.
We assume a steady-state setting in which the following  \textit{mean-square system cost}   is well defined, and finite a.s.,
\begin{equation}
J^2 =  \lim_{T\to\infty}\frac{1}{T} \int_0^T | \UA_a(t) |^2\, dt
\label{e:Ja}
\end{equation}
This has a representation via spectral theory of stationary processes~\cite{cai88}.  

The power spectral density $P_D$ for $\bfmD$ can be estimated by fitting observed data to a linear system driven by white noise \cite{cai88}.  As done previously in \cite{chebusmey14},  this approach was used to obtain an estimate of $P_D$ for the BPA balancing reserves. The resulting power spectral density for a stationary version of $\bfmUA_a$ is given by 
\[
P_{\UA_a}(\omega) = P_D(\omega) \Big|\frac{\UA_a}{D}(j\omega)\Bigr|^2
\]
where the transfer function is defined in \eqref{e:UaDsec3}.
The mean-square system cost is expressed,
\begin{equation}
\label{e:J}
J^2 = \frac{1}{2\pi}\int_{-\infty}^{\infty}  P_{\UA_a}(\omega)\,  d\omega\,  .
\end{equation}

\Fig{f:H2} shows two plots of  $J$ as a function of $\rho$   for several values of $\omega_{co}$. For comparison,  in  each plot, we include the cost plot for $\omega_{co}=0.013$~rad/s and an ``ideal TCL'' in which $G_{tcl}(s)\equiv 1$.

In the lead design, 
the cost is near its minimum when $\rho = 1$ and $\omega_{co} = 0.003$~rad/s. In other words, 
with this lead design,
we can minimize the system cost by including all the available TCL resources provided they service regulation signals with frequencies below 0.003 rad/s. 

The bandwidth of TCL service can be increased significantly with more aggressive local control. For the inverse design, the cost is at its lowest for $\omega_{co}=0.013$~rad/s. The  performance  nearly matches the ideal performance in this case.

\subsection{Disturbance rejection}
\label{s:track}


 The disturbance-rejection performance of the system was evaluated using Simulink. 
 Multiple experiments were conducted to evaluate performance using several different designs.  
 
 In each case, $\rho$ was set to $1$ in the definition of the actuator block $H$ (see  \eqref{e:H}).
The remaining components of the block diagram shown in \Fig{f:CAv2} are as specified in this section.

In summary,
\begin{romannum}
\item When lead control is absent, the closed loop system is unstable for $\omega_{co} > 0.003$~rad/s, which is the natural frequency of $G_{tcl}$.

\item  For the lead design,  disturbance rejection is very good for $\omega_{co} \leq 0.005$~rad/s; the system is unstable for $\omega_{co} \geq 0.009$~rad/s.

\Fig{f:TrackingLead007}  shows results for  $\omega_{co} = 0.007$~rad/s.
   \notes{always assumed?  But elsewhere I thought we said one week!
   \\
   and with $\bfmD $ equal to the balancing reserves at BPA during a typical day (June 6, 2015).   }
   The oscillation is due to a strong resonance in the closed loop system --- this is caused by phase lag in $H_{tcl}$.

\item
Disturbance rejection is improved using the inverse design 
for the TCLs.   Results obtained with $\omega_{co} = 0.013$~rad/s are shown in  
\Fig{f:TrackingInverseFilter013}.  
The disturbance rejection observed here is the best among all  experiments. 

\end{romannum}

\section{Conclusions}
\label{s:conc}

With appropriate filtering and local control, Demand Dispatch can provide excellent ancillary service in low-to-mid frequency ranges, even without two-way communication.  While there is some cost to install hardware on appliances that can receive a signal from a balancing authority,  in the long run this will be far less expensive than batteries.

The grid operator does require some bounds on the behavior of loads.  In particular, 
gain uncertainty is a potential concern, especially in frequency ranges corresponding to PJM's RegD signal and above.   It may be valuable for 
each load to broadcast its state occasionally -- perhaps just once per hour --  so that capacity can be estimated at the grid level \cite{chebusmey15}.
  
The best communication architecture for higher frequencies (corresponding to primary reserves) remains a research frontier.

\bibliographystyle{abbrv}   {\small

\def\cprime{$'$}\def\cprime{$'$}


\begin{thebibliography}{10}

\bibitem{BPA}
{BPA} balancing authority.
\newblock Online, \url{tinyurl.com/BPAgenload} and
  \url{tinyurl.com/BPAbalancing}.

\bibitem{barbusmey14}
P.~Barooah, A.~Bu\v{s}i\'{c}, and S.~Meyn.
\newblock Spectral decomposition of demand-side flexibility for reliable
  ancillary services in a smart grid.
\newblock In {\em Proc. {48th Annual Hawaii International Conference on System
  Sciences (HICSS)}}, pp.~2700--2709, Kauai, Hawaii, 2015.

\bibitem{MISO13}
K.~E. Bennett.
\newblock Status of {MISO's} regulation market since implementation of the
  performance-based revisions as required by {Order No. 755. Midcontinent
  Independent System Operator, Inc. -- Informational Report}.
\newblock Report prepared for the {FERC}. Docket \# ER12-1664, Midwest ISO,
  Senior Corporate Counsel, 2013.

\bibitem{broThesis09}
K.~M. Brokish.
\newblock Adaptive load control of microgrids with non-dispatchable generation.
\newblock Master's thesis, Massachusetts Institute of Technology, 2009.

\bibitem{bro10}
A.~Brooks, E.~Lu, D.~Reicher, C.~Spirakis, and B.~Weihl.
\newblock Demand dispatch.
\newblock {\em Power and Energy Magazine, IEEE}, 8(3):20--29, 2010.

\bibitem{cai88}
P.~E. Caines.
\newblock {\em {Linear Stochastic Systems}}.
\newblock John Wiley \& Sons, New York, 1988.

\bibitem{chaghaeri14}
H.~Chamorro, M.~Ghandhari, and R.~Eriksson.
\newblock Influence of the increasing non-synchronous generation on small
  signal stability.
\newblock In {\em {IEEE PES General Meeting}}, pp.~1--5, July 2014.

\bibitem{chabalsha12}
H.~Chavez, R.~Baldick, and S.~Sharma.
\newblock Regulation adequacy analysis under high wind penetration scenarios in
  {ERCOT} nodal.
\newblock {\em IEEE Trans. on Sustainable Energy}, 3(4):743--750, Oct 2012.

\bibitem{chebusmey14}
Y.~Chen, A.~Bu\v{s}i\'{c}, and S.~Meyn.
\newblock Individual risk in mean-field control models for decentralized
  control, with application to automated demand response.
\newblock In {\em {Proc. of the 53rd IEEE Conference on Decision and Control}},
   pp.~6425--6432, Dec. 2014.

\bibitem{chebusmey15}
Y.~Chen, A.~Bu\v{s}i\'{c}, and S.~Meyn.
\newblock State estimation and {Mean-Field Control} with application to demand
  dispatch.
\newblock In {\em {54th IEEE Conference on Decision and Control} (to appear)
  and arxiv.org/abs/1504.00088v1}, Dec. 2015.

\bibitem{haomidbarey12}
H.~Hao, T.~Middelkoop, P.~Barooah, and S.~Meyn.
\newblock How demand response from commercial buildings will provide the
  regulation needs of the grid.
\newblock In {\em 50th {Allerton} Conference on Communication, Control, and
  Computing}, pp.~1908--1913, 2012.

\bibitem{kir04}
B.~J. Kirby.
\newblock Frequency regulation basics and trends.
\newblock Report prepared for the {US DoE} -- ORNL/TM-2004/291.

\bibitem{kun94}
P.~Kundur.
\newblock {\em Power system stability and control}, volume~7 of {\em {EPRI
  power system engineering}}.
\newblock McGraw-Hill New York, 1994.

\bibitem{linbarmeymid15}
Y.~Lin, P.~Barooah, S.~Meyn, and T.~Middelkoop.
\newblock Experimental evaluation of frequency regulation from commercial
  building {HVAC} systems.
\newblock {\em IEEE Trans. on Smart Grid}, 6(2):776--783, 2015.

\bibitem{malcho88}
R.~Malham\'e and C.-Y. Chong.
\newblock On the statistical properties of a cyclic diffusion process arising
  in the modeling of thermostat-controlled electric power system loads.
\newblock {\em SIAM J. Appl. Math.}, 48(2):pp. 465--480, 1988.



\bibitem{johThesis12}
J.~Mathieu.
\newblock {\em Modeling, Analysis, and Control of Demand Response Resources}.
\newblock PhD thesis, University of California at Berkeley, 2012.

\bibitem{meybarbusyueehr14}
S.~Meyn, P.~Barooah, A.~Bu\v{s}i\'{c}, Y.~Chen, and J.~Ehren.
\newblock Ancillary service to the grid using intelligent deferrable loads.
\newblock 
{\em To appear, IEEE Trans.\ on
  Auto. Control}, 2015.

\bibitem{milshapajdaq14}
N.~Miller, M.~Shao, S.~Pajic, and R.~D'Aquila.
\newblock Western wind and solar integration study phase 3--frequency response
  and transient stability.
\newblock  NREL Technical report,
  Golden, CO., 2014.
 
 
\bibitem{FERC755}
R.~Pedroncelli.
\newblock {Frequency Regulation Compensation in the Organized Wholesale Power
  Markets -- FERC 755}.
\newblock {FERC Docket Nos. RM11-7-000 and AD10-11-000; Order No. 755} --
  Online, \url{http://tinyurl.com/FERC755}, October 20 2011.

\bibitem{peybal12}
M.~Peydayesh and R.~Baldick.
\newblock The effects of very fast response to frequency fluctuation.
\newblock In {\em {Proceedings of the 31st USAEE/IAEE North American
  Conference}}. {International Association for Energy Economics}, 2012.

\bibitem{pjm-man15}
C.~Pilong.
\newblock {PJM Manual 12: Balancing Operations}.
\newblock Online \url{tinyurl.com/pjm-man15}, April 2015.



\bibitem{schFAPER80}
F.~Schweppe, R.~Tabors, J.~Kirtley, H.~Outhred, F.~Pickel, and A.~Cox.
\newblock Homeostatic utility control.
\newblock {\em IEEE Trans. on Power Apparatus and Systems}, PAS-99(3):1151--1163, May 1980.

\bibitem{xiasubrefrecarschola14}
Y.~Xiao, Q.~Su, S.~Bresler, S.~Frederick, R.~Carroll, J.~R. Schmitt, and
  M.~Olaleye.
\newblock Performance-based regulation model in {PJM} wholesale markets.
\newblock In {\em PES General Meeting| Conference \&\ Exposition, 2014 IEEE},
  pp.~1--5. IEEE, 2014.

\end{thebibliography}
}

\null  

  \end{document}